\documentclass{emulateapj}
\usepackage{graphicx,natbib,amsmath,soul,xspace,rotating} 
\usepackage{CJKutf8}

\newcommand{\sersic}{S\'{e}rsic}

\newcommand{\Lsun}{$L_{\odot}$}
\newcommand{\Msun}{$M_{\odot}$}
\newcommand{\LIR}{$L_{\rm IR}$}  
\newcommand{\mum}{$\mu m$}
\newcommand{\Mstar}{$M_{*}$}

\newcommand{\Ha}{H$\alpha$}

\newcommand{\scuba}{\textsc{Scuba\,2}\xspace}

\begin{document}

\title{Large scale structure around a $z=2.1$ cluster}

\author{Chao-Ling Hung \begin{CJK*}{UTF8}{bsmi}(洪肇伶)\end{CJK*}\altaffilmark{1,2}}
\author{Caitlin M. Casey\altaffilmark{1}}
\author{Yi-Kuan Chiang\altaffilmark{1}}
\author{Peter L. Capak\altaffilmark{3,4}}
\author{Michael J. Cowley\altaffilmark{5,6}}
\author{Behnam Darvish\altaffilmark{4}}
\author{Glenn G. Kacprzak\altaffilmark{7}}
\author{K. Kova\v{c}\altaffilmark{8}}
\author{Simon J. Lilly\altaffilmark{8}}
\author{Themiya Nanayakkara\altaffilmark{7}}
\author{Lee R. Spitler\altaffilmark{5,6}}
\author{Kim-Vy H. Tran\altaffilmark{9}}
\author{Tiantian Yuan\altaffilmark{10}}

\affil{\altaffilmark{1} Department of Astronomy, the University of Texas at Austin, 2515 Speedway Blvd., Austin, TX 78712, USA}
\affil{\altaffilmark{2} Harlan J. Smith Fellow}
\affil{\altaffilmark{3} Infrared Processing and Analysis Center, California Institute of Technology, Pasadena, CA 91125, USA}
\affil{\altaffilmark{4} Cahill Center for Astronomy and Astrophysics, California Institute of Technology, Pasadena, CA 91125, USA}
\affil{\altaffilmark{5} Department of Physics and Astronomy, Macquarie University, NSW 2109, Australia}
\affil{\altaffilmark{6} Australian Astronomical Observatory, PO Box 915, North Ryde, NSW 1670, Australia}
\affil{\altaffilmark{7} Swinburne University of Technology, Hawthorn, VIC 3122, Australia}
\affil{\altaffilmark{8} Institute for Astronomy, Department of Physics, ETH Zurich, CH-8093 Zurich, Switzerland}
\affil{\altaffilmark{9} George P. and Cynthia W. Mitchell Institute for Fundamental Physics and Astronomy, Department of Physics \& Astronomy, Texas A\&M University, College Station, TX 77843}
\affil{\altaffilmark{10} Research School of Astronomy and Astrophysics, The Australian National University, Cotter Road, Weston Creek, ACT2611, Australia}


\begin{abstract}
The most prodigious starburst galaxies are absent in massive galaxy clusters today, but their connection with large scale environments is less clear at $z\gtrsim2$.
We present a search of large scale structure around a galaxy cluster core at $z=2.095$ using a set of spectroscopically confirmed galaxies.
We find that both color-selected star-forming galaxies (SFGs) and dusty star-forming galaxies (DSFGs) show significant overdensities around the $z=2.095$ cluster.
A total of 8 DSFGs (including 3 X-ray luminous active galactic nuclei, AGNs) and 34 SFGs are found within a 10$\arcmin$ radius (corresponds to $\sim$15 cMpc at $z\sim2.1$) from the cluster center and within a redshift range of $\Delta z=0.02$, which leads to galaxy overdensities of $\delta_{\rm DSFG}\sim12.3$ and $\delta_{\rm SFG}\sim2.8$.
The cluster core and the extended DSFG- and SFG-rich structure together demonstrate an active cluster formation phase, in which the cluster is accreting a significant amount of material from large scale structure while the more mature core may begin to virialize.  
Our finding of this DSFG-rich structure, along with a number of other protoclusters with excess DSFGs and AGNs found to date, suggest that the overdensities of these rare sources indeed trace significant mass overdensities.
However, it remains puzzling how these intense star formers are triggered concurrently. 
Although an increased probability of galaxy interactions and/or enhanced gas supply can trigger the excess of DSFGs, our stacking analysis based on 850 $\mu$m images and morphological analysis based on rest-frame optical imaging do not show such enhancements of merger fraction and gas content in this structure.


\end{abstract}


\section{Introduction}
Large-scale structure plays a critical role on the evolution and assembly of galaxies.
Red and massive elliptical galaxies are preferentially located in high density environments today \citep[e.g.,][]{Dressler1980,Postman1984,Hogg2004}, where star formation is strongly suppressed \citep[e.g.,][]{Hashimoto1998,Gomez2003,Goto2005}.
In a hierarchical structure formation paradigm, these correlations with local environments should become less significant and even overturn at earlier epochs \citep[e.g.,][]{Hopkins2008}.
Although some studies have observed the reversal of the star formation-density relation at $z\sim1$ \citep{Elbaz2007,Cooper2008}, exactly when and where these correlations with local environments are established remains highly uncertain \citep[e.g.,][]{Patel2009,Tran2010,Brodwin2013,Smail2014,Darvish2014,Darvish2016}.
In fact, the existence of mature cluster cores and evidence of early mass assembly of brightest cluster galaxies at $z\sim1-2$ suggest that the formation of these rare objects takes place rapidly at even higher redshifts \citep[e.g,][]{Collins2009,Gobat2011,Zeimann2012}. 
It is thus critical to probe beyond $z\sim2$ so we can observe the formation of massive galaxy clusters and directly infer the early environmental influences on galaxy evolution.

Common probes of galaxy clusters \citep[e.g., X-ray emission from hot intracluster gas, red galaxy sequence, inverse Compton scatter of the Cosmic Microwave Background photons off the hot intracluster medium;][]{Allen2011} reach their limits at $z\sim2$ due to the combination of survey sensitivity and the less-evolved nature of large scale structure.
Identification of protoclusters at $z\gtrsim2$ thus often rely on significant overdensities of galaxies on the sky and in the redshift space \citep[e.g.,][]{Steidel1998,Miley2004}.
One successful strategy to identify high$-z$ protoclusters is to place the survey area around highly biased galaxies like radio galaxies and quasars since they are likely progenitors of massive elliptical galaxies in the core of present-day galaxy clusters \citep[e.g.,][]{Kurk2000,Miley2004,Venemans2007,Utsumi2010,Capak2011,Hatch2011,Hayashi2012,Koyama2013a,Wylezalek2013}.
Meanwhile, some protoclusters have also been discovered serendipitously via narrow-band imaging or spectroscopic surveys \citep[e.g.,][]{Steidel1998,Shimasaku2003,Toshikawa2012,LeeKS2014}.
In either case, the $z\gtrsim2$ protoclusters are often traced by optically-selected galaxies like (spectroscopically-confirmed) Lyman Break galaxies (LBGs), Lyman-$\alpha$ emitters (LAEs), and/or \Ha\ emitters (HAEs).

Recent advent of large submillimeter surveys have enabled an alternative window to search for $z\gtrsim2$ protoclusters.
Clustering analysis of submillimeter galaxies \citep[SMGs,][]{Blain2002} show that SMGs are located in massive halos at $z\sim2$ \citep[e.g.,][]{Blain2004,Hickox2012}, yet it remains highly debated whether SMGs indeed trace the densest environments at $z=0$ \citep{Capak2011,Carilli2011,Miller2015,Cowley2015}.
Observations of SMG overdensities have led to contradictory interpretations.
\citet{Casey2015} discover a highly unusual filamentary structure at $z=2.47$ containing 7 dusty star-forming galaxies \citep*[DSFGs with star formation rate (SFR) $\gtrsim$200 \Msun yr$^{-1}$;][]{Casey2014}, 5 Active Galactic Nuclei (AGNs with $L_X\gtrsim$ 10$^{43.5}$ erg s$^{-1}$), and 33 other spectroscopically confirmed LBGs within a comoving volume of 15,000 Mpc$^3$.
Such overdensities in DSFGs ($\delta_{\rm DSFG}=10$) and AGNs ($\delta_{\rm AGN}=35$) are unlikely to be a result of survey incompleteness or biases, and \citet{Casey2015} argue that this structure is a possible progenitor of the present-day Coma-like cluster.
On the other hand, \citet{Blain2004} and \citet{Chapman2009} identify an association of SMGs at $z=1.99$ in the GOODS-N field, in which \citet{Chapman2009} find a strong overdensity of SMGs ($\delta_{\rm SMG}=10$) but only a moderate overdensity of UV-selected galaxies ($\delta_{\rm UV}=2.5$).
Based on the linear theory of spherical collapse, \citet{Chapman2009} conclude that the structure traced by UV-selected galaxies will not collapse by $z=0$\footnote{However, Casey (2016) argue that when comparing with cosmological simulations in \citet{Chiang2013}, this structure is among the top 30\% of structures that will collapse by $z=0$.}, and they attribute the contradictory conclusions based on SMG overdensities as a result of an even larger galaxy bias of SMGs than they assumed.

Protoclusters that are identified through both submillimeter and optical windows can provide crucial insights to understand the nature of DSFG-rich structures and their differences with protoclusters traced by optically-selected galaxies.
\citet[][also see \citealp{Umehata2015}]{Tamura2009} present an imaging survey of 1.1 mm emission in SSA 22 \citep[a $z=3.09$ protocluster,][]{Steidel1998}, and they find that the population of SMGs is enhanced near the protocluster core and there is a spatial correlation between SMGs and LAEs. 
In MRC1138$-$262 \citep[a $z=2.16$ protocluster,][]{Kurk2000}, \citet{Dannerbauer2014} find an excess of SMGs in the protocluster yet the concentration of SMGs does not coincide with the central radio galaxy.
Such offset between dusty starburst population and the densest regions of protocluster is also seen in a $z=1.62$ structure \citep{Smail2014}. 
Does the spatial distribution of dusty starbursts represent the evolutionary status of protoclusters?
Then what are the implications when using DSFGs as tracers of large scale structure?
A systematic search of DSFG populations in the known protoclusters is necessary to shed light on these questions.

In this paper, we present a search of DSFGs and large scale structure around a bona-fide galaxy cluster at $z=2.095$ \citep[][hereafter Y14]{Yuan2014}. 
We provide an overview of this $z=2.095$ cluster in Section 1.1 and we detail our analysis in Section 2.
In Section 3, 4, and 5, we present our results on the large scale structure found around the $z=2.095$ cluster, the environmental dependence of galaxy properties, and a detailed scrutiny of DSFGs in the structure.
We discuss the implications of our results and provide a brief summary in Section 6. 
Throughout this paper, we adopt a $\Lambda$CDM cosmology with $H_0=70$ km s$^{-1}$ Mpc$^{-1}$, $\Omega_{M}=0.3$ and $\Omega_{\Lambda}=0.7$.

\subsection{A $z=2.095$ cluster in the COSMOS field}
\citet{Spitler2012} presented a candidate protocluster at $z=2.2$\footnote{This structure is located at $z=2.1$ based on updated photometric measurements (L. Spitler; private communications).} located in the central region of the COSMOS field \citep[][]{Scoville2007}. 
This discovery is based on an overdensity of red galaxies selected via medium-bandwidth near-IR imaging from ZFOURGE \citep[][Straatman et al. {\it submitted}]{Tomczak2014}.
No significant diffuse X-ray emission is detected at the location of this structure \citep{Spitler2012}.
The galaxy density maps constructed based on photometric-redshifts from \citet{Muzzin2013} also recover the same structure \citep[located at $z=2.07$,][]{Chiang2014}.
Y14 presented the spectroscopic confirmation of this structure, in which they identified 57 confirmed cluster members with a median redshift of $z=2.095$ (hereafter the ZFIRE cluster).
The ZFIRE cluster identified by Y14 spans a $\sim$12\arcmin$\times$12\arcmin\ region on the sky and have a velocity dispersion of $\sigma\sim$ 552 km\,s$^{-1}$.  
Based on the comparison with cosmological simulations, Y14 concluded that the ZFIRE structure may evolve to a Virgo-like cluster ($M_{\rm vir} \sim 10^{14.4}$\Msun) at $z=0$.
There have been several ongoing efforts to characterize the physical properties of the member galaxies in the ZFIRE cluster.
The mass-metallicity relation in this cluster is consistent to the field \citep{Kacprzak2015}, and detailed studies of rest-frame optical line ratios toward 13 cluster members also find no significant differences in ISM properties of galaxies in the cluster and field galaxies at the same redshift \citep{Kewley2015}.

\section{Data and Analysis}
To search for the populations of DSFGs within and near the ZFIRE cluster, we draw a sample of spectroscopically confirmed galaxies at $2.07\leq z \leq 2.12$ from a number of redshift surveys. 
We note that the ZFIRE cluster members in Y14 span a redshift range of $2.08\lesssim z \lesssim2.11$, and here we enlarge this redshift range by 0.01 to search for galaxies in the large structure that may also be associated with the cluster core. However, all of the cluster members defined based on the friends-of-friends analysis (Section 3.1) fall within a similar redshift range as the ZFIRE cluster members.
The redshift surveys included in this work are listed below:
\begin{enumerate}
\item The ZFIRE survey from Y14 and Nanayakkara et al. (2016, {\it submitted}). The targets were originally selected based on photometric redshifts from \citet{Spitler2012} as part of the ZFOURGE survey and observed with Keck I Multi-Object Spectrometer For Infra-Red Exploration (MOSFIRE). The ZFIRE survey achieves a detection limit (S/N$\sim$5) of objects with $Ks\sim25$. 
\item The redshift survey of {\it Herschel} SPIRE-selected and \scuba-selected sources from \citet{Casey2012a} and Casey et al. (2016, \textit{in prep.}). The submillimeter-bright galaxies were observed with Keck I Low Resolution Imaging Spectrometer (LRIS), Keck II DEep Imaging Multi-Object  Spectrograph (DEIMOS), and MOSFIRE.
\item The zCOSMOS-deep redshift survey from \citet{Lilly2007} and Lilly et al. (2016, \textit{in prep.}). The targeted galaxies were selected based on $BzK$ criteria \citep{Daddi2004} and UGR criteria \citep{Steidel1996} with $K_{\rm AB}\lesssim23.5$ \citep{Lilly2007,Diener2013}, in which the sample  yields a set of star-forming galaxies that are mostly at $1.3\lesssim z \lesssim 3$. 
We exclude sources with insecure redshift measurements that are also inconsistent with photometric redshifts estimates (flag 1.1 and 2.2 defined by \citet{Lilly2009}). 
These star-forming galaxies were observed with VLT VIMOS spectrograph.
\item Additional public MOSDEF and VUDS redshift catalogs are also included our analysis \citep{Kriek2015,Tasca2016}. Both of these redshift surveys cover the CANDELS-COSMOS field, and thus the ZFIRE cluster.
\end{enumerate}

We obtain the full UV-near-IR multiwavelength data set of spectroscopically-confirmed sources through the COSMOS photometric redshift catalogs \citep{Capak2007,Ilbert2009,McCracken2010}.
When available, mid-far IR photometry are obtained from the {\it Spitzer}-COSMOS survey \citep{Sanders2007}, the PACS Evolutionary Probe program \citep[PEP;][]{Lutz2011}, the Herschel Multitiered Extragalactic Survey \citep[HerMES][]{Oliver2012}, and the \scuba\ 450 \mum\ and 850 \mum\ survey from \citet{Casey2013}.
The MIPS 24 \mum\ through SPIRE 500 \mum\ catalog is compiled in \citet{LeeN2013}.
The spectroscopically-confirmed sources from all redshift catalogs were also cross-correlated with the {\it Chandra}-COSMOS catalog \citep{Elvis2009,Civano2012} to search for X-ray luminous AGNs in and near the ZFIRE cluster.
In this study, we focus only on X-ray selected AGNs, and we refer the reader to \citet{Cowley2016} for the population of mid-IR selected and radio-selected AGNs in the ZFIRE cluster.

\begin{figure*}
\centering
  \includegraphics[width=0.9\textwidth]{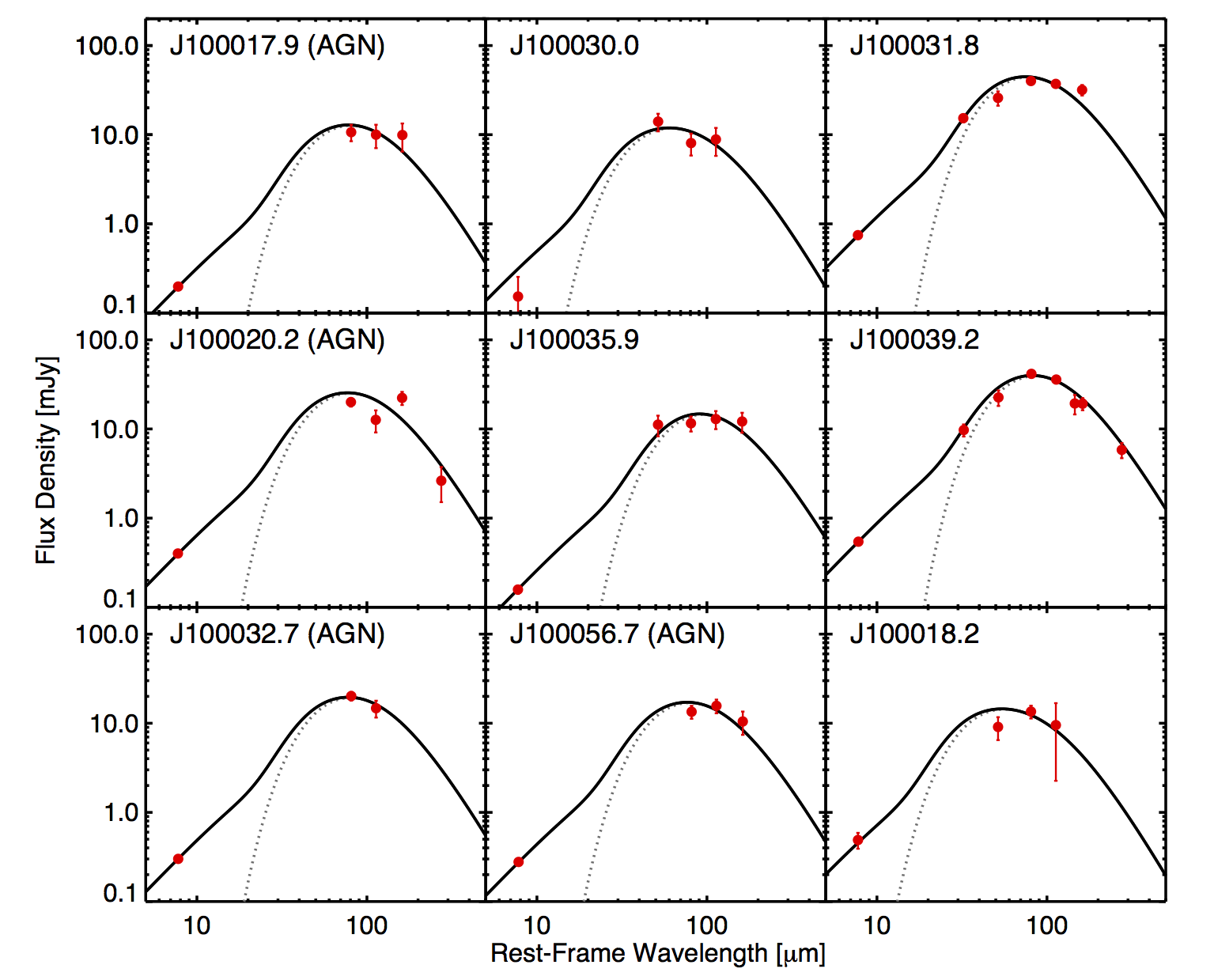} 
\caption{Rest-frame far-infrared SED of 9 DSFGs (including 4 X-ray luminous AGNs) within a 10\arcmin\ circle from the center of the ZFIRE cluster and a redshift range of $2.07\leq z \leq 2.12$.  
The MIPS, PACS, SPIRE, and \scuba\ photometric data points and uncertainties are shown in red. 
The best-fit SED using a coupled graybody and mid-IR power law \citep{Casey2012} is shown in a black solid line, with the underlying graybody shown as a dotted line.
} 
\label{fig:sedpanel}
\end{figure*}

We require potential DSFGs to have 24 \mum\ detections and are detected (S/N$\geq$3) in at least two additional bands from PACS 100 \mum, 160 \mum, SPIRE 250 \mum, 350 \mum, 500 \mum, or \scuba\ 450 \mum, 850 \mum.
We measure the far-infrared luminosity (\LIR $\equiv L_{8-1000\mu m}$ in the object rest-frame), dust temperature, and dust mass of these sources by fitting their FIR spectral energy distributions (SEDs) to a coupled modified gray body and mid-IR power law \citep{Casey2012}.
We assume a dust emissivity ($\beta$) of 1.5, and the results do not change significantly with $\beta=1.5-2.0$.
Further, we assumed the slope of mid-infrared power law ($\alpha$) of 2.
In the cases where more than three data points are available at rest-frame wavelengths shorter than $\sim70$ \mum, we do not see significant differences when leaving $\alpha$ as a free fitting parameter.
Within a 10\arcmin\ circle from the center of the ZFIRE cluster\footnote{Here we use the median position of ZFIRE cluster members as the cluster center (the same position as the center used in Y14). Although the ZFIRE cluster shows several sub-structures and the position of cluster center is highly uncertain, our search of DSFG overdensity is insensitive to the chosen reference point in the ZFIRE cluster.} and a redshift range of $2.07\leq z \leq 2.12$, we find a total of 9 sources with \LIR\ $\geq$ $10^{12}$ \Lsun\footnote{This definition is the same as the selection of ultraluminous infrared galaxies based on rest-frame far-infrared emission \citep*[e.g.,][]{Sanders1996,Casey2014}. All but one of these {\it Herschel}-selected DSFGs also fall in the dusty star-forming region defined based on the rest-frame UVJ color selection \citep{Wuyts2007,Spitler2014}.}.
Four X-ray luminous AGNs are found within the same search area, and all of them are DSFGs.
The IR and X-ray properties of the 9 DSFGs (including 4 X-ray luminous AGNs) identified within a 10\arcmin\ circle from the center of the ZFIRE cluster are summarized in Table~\ref{tab:dsfg}, and the best-fit FIR SEDs are shown in Figure~\ref{fig:sedpanel}.

\begin{sidewaystable*}
\centering
 \caption{Properties of DSFGs associated to the ZFIRE cluster}
 \label{tab:dsfg}
\begin{tabular}{@{}lcccccccccccc}
 \hline
 \hline
Name& $z$ & Redshift & $S_{24}$ & $S_{100}$ & $S_{160}$ & $S_{250}$ & $S_{350}$ & $S_{450}$ & $S_{500}$& $S_{850}$ &\LIR & $L_X$  \\
        &       & Survey  &   ($\mu$Jy)  & (mJy)  & (mJy)  & (mJy)  & (mJy) & (mJy)  & (mJy)  & (mJy)  & ($10^{12}$ \Lsun) & ($10^{43}$ erg\,s$^{-1}$)    \\
 \hline
J100017.9$+$021807.2&  2.094 & (1)\footnote{Redshift surveys: (1) ZFIRE, (2) SPIRE and \scuba-bright sources, (3) zCOSMOS, and (4) MOSDEF.} \Ha\footnote{The main spectral line used to determine redshift.} & 198$\pm$14 & ... & ... & 10.7$\pm$2.2\footnote{Uncertainties of SPIRE photometry refer to the quadrature sum of instrumental and confusion noise \citep{Smith2012}.} & 10.0$\pm$2.9 & ... & 9.9$\pm$3.4 & ... & 2.58$^{+1.05}_{-0.75}$ & 20.2$\pm$2.8 \\
J100030.0$+$021413.1&  2.098 & (1) \Ha\ & 153$\pm$101 & ... & 14.0$\pm$3.1 & 8.1$\pm$2.2 & 8.8$\pm$3.1& ... & ... & ... & 3.19$^{+3.12}_{-1.58}$  & ...  \\
J100031.8$+$021242.7&  2.104 & (1) \Ha\ & 746$\pm$18 & 15.3$\pm$1.6 & 25.9$\pm$4.7 & 40.0$\pm$2.2 & 37.1$\pm$3.2 & ... & 31.8$\pm$4.3 & ... & 9.47$^{+0.96}_{-0.87}$ & ... \\

J100020.2$+$021725.7&  2.104 & (2) \Ha\ & 401$\pm16$ & ... & ... & 20.0$\pm$2.2 & 12.6$\pm$3.5 & ... & 22.3$\pm$3.7 & 2.6$\pm$1.1 & 5.23$^{+1.06}_{-0.88}$ & 23.9$\pm$2.7\\
J100035.9$+$021128.1&  2.103 & (2) Ly$\alpha$ & 158$\pm$17 & ... & 11.2$\pm$2.9 & 11.6$\pm$2.2 & 12.9$\pm$3.0 & ... & 12.1$\pm$3.1 & ... & 2.50$^{+0.97}_{-0.70}$ &  ... \\  
J100039.2$+$022220.9&  2.085 & (2) \Ha\ & 544$\pm$17 & 9.7$\pm$1.5 & 22.6$\pm$4.5 & 41.5$\pm$2.2 & 35.8$\pm$2.7 & 19.3$\pm$4.8 & 19.2$\pm$3.0 & 5.8$\pm$1.1 & 7.47$^{+0.76}_{-0.69}$ & ... \\

J100032.7$+$021331.1&  2.091 & (3) Ly$\alpha$ & 301$\pm$15 & ... & ... & 20.2$\pm$2.2 & 14.7$\pm$3.2 & ... & ... & ... & 3.94$^{+1.10}_{-0.86}$  & 5.6$\pm$1.5\\
J100056.7$+$021720.9&  2.076 & (3) Ly$\alpha$ & 278$\pm$17 & ... & ... & 13.5$\pm$2.2 & 15.7$\pm$2.7 & ... & 10.5$\pm$3.1 & ... & 3.47$^{+0.94}_{-0.74}$  & 20.5$\pm$2.4 \\

J100018.2$+$021842.6&  2.102 & (4) \Ha\ & 490$\pm$99 & ... & 9.1$\pm$2.6 & 13.5$\pm$2.2 & 9.5$\pm$7.3 & ... & ... & ... & 4.28$^{+1.83}_{-1.28}$ & ... \\

\hline

\end{tabular}
\end{sidewaystable*}

\begin{figure}
  \includegraphics[width=0.5\textwidth]{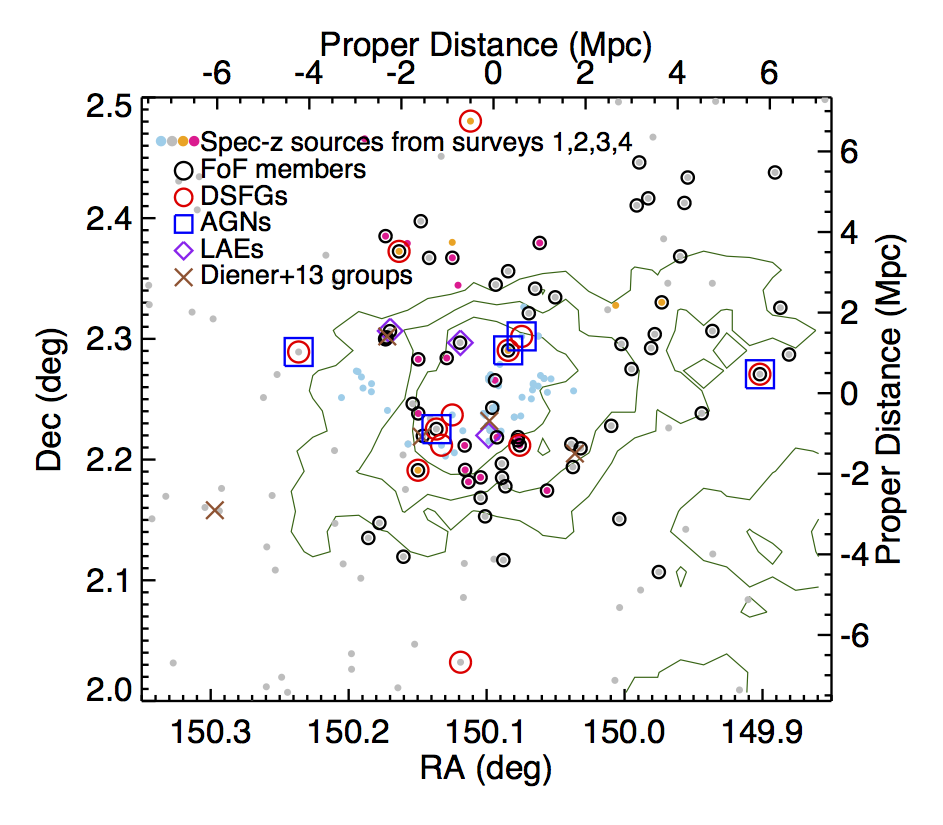} 
  \includegraphics[width=0.5\textwidth]{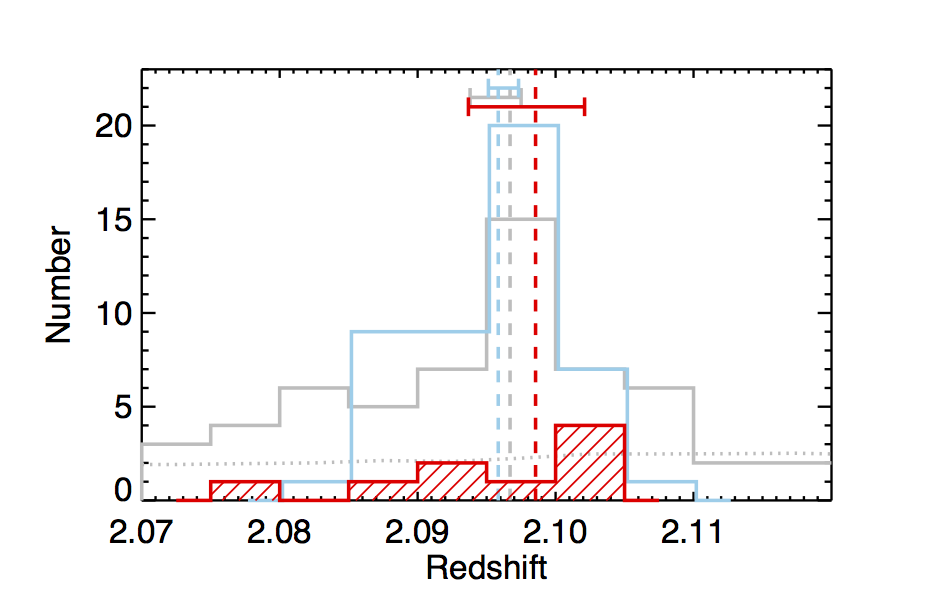} 
\caption{(\textit{Top}) Spatial distributions of all spectroscopically confirmed galaxies at the redshift range of $2.07\leq z \leq 2.12$ from four redshift surveys (light blue dots: ZFIRE survey, orange dots: SPIRE/\scuba\ survey, gray dots: zCOSMOS survey, pink dots: MOSDEF/VUDS surveys).
The right and top axes label the proper distance at $z=2.1$ with respect to the cluster center.
Black circles indicate sources that are linked to the ZFIRE cluster members with a linking length of 2 Mpc.
Red circles show galaxies with \LIR\ $\geq10^{12}$ \Lsun, and blue boxes indicate X-ray sources.
Purple diamonds show the positions of three LAEs from \citet{Chiang2015}.
Brown crosses mark the group candidates identified by \citet{Diener2013}.
Background dark green contours outline the $z=2.07$ structure identified based on photometric redshifts ($\delta_{\rm gal}$=0.2,0.5,0.8 from Figure 2 in \citet{Chiang2014}).
\textit{(Bottom)} Redshift distribution of the zCOSMOS SFGs (gray solid line), ZFIRE SFGs (sky blue solid line), and DSFGs (red shaded area) at $2.07\leq z \leq 2.12$ within a 10\arcmin radius from the ZFIRE cluster center.
The dashed lines indicate the median redshifts of each sample, and the error bars represent the bootstrapped uncertainties.
The gray dotted line represents the expected number of SFGs from the zCOSMOS survey.
} 
\label{fig:dist_2dsky}
\end{figure}

\section{Large Scale Structure around the ZFIRE cluster}

\subsection{Distribution of star-forming galaxies and rare sources}
Figure~\ref{fig:dist_2dsky} shows the spatial distribution of all spectroscopically confirmed sources at $2.07\leq z \leq 2.12$ ($\Delta z=0.05$ corresponds to a velocity range $\Delta v\sim4800$ km s$^{-1}$) and their distribution in the redshift space.
Within a 10\arcmin\ (corresponds to a proper distance of $\sim5$ Mpc) radius from the ZFIRE cluster center, the redshift distribution of star-forming galaxies (SFGs) from the zCOSMOS survey peaks at similar redshift (median $z=2.097^{+0.001}_{-0.003}$, the error represents the bootstrapped uncertainties) as the ZFIRE cluster members. 
We perform the friends-of-friends (FoF) analysis \citep{Huchra1982} to determine which galaxies from the zCOSMOS, SPIRE/\scuba\ surveys, and MOSDEF/VUDS can be linked to the ZFIRE cluster members (from redshift survey (1); blue dots in Figure~\ref{fig:dist_2dsky}) with a linking length of 2 Mpc, a scale that accommodates the large spatial extent of un-virialized structure at $z>2$ \citep[e.g.][]{Chiang2013,Muldrew2015}.
The structure identified from the FoF algorithm (black circles in Figure~\ref{fig:dist_2dsky}) spans a comoving volume of $\sim25,000$ Mpc$^{3}$, and its spatial extent is consistent with the $z=2.07$ structure identified based on photometric redshifts \citep{Chiang2014}.
This structure also covers three candidate galaxy groups selected based on zCOSMOS survey \citep{Diener2013}.

The rare sources (DSFGs and AGNs) identified via all redshift surveys span an area of $\sim20\arcmin \times 20\arcmin$ on the sky.
For the 9 DSFGs (4 of them are X-ray luminous AGNs) that fall within a 10\arcmin\ radius from the ZFIRE cluster center, they have comparable median redshift ($z=2.099^{+0.004}_{-0.005}$) as the ZFIRE cluster members and zCOSMOS SFGs (we note that here 3 DSFGs are drawn from ZFIRE and 2 DSFGs are drawn from zCOSMOS surveys). 
If we define potential cluster members using the FoF analysis instead of a fixed 10\arcmin\ aperture, then there are also 9 DSFGs (4 of them are X-ray luminous AGNs) included in the large scale structure.
Such an excess of DSFGs and AGNs within and around the ZFIRE clusters is comparable to several DSFGs and AGNs rich structures at $z\gtrsim1.5$ \citep[][see a compilation by \citealp{Casey2016}]{Chapman2009,Tamura2009,Digby-North2010,Dannerbauer2014,Smail2014,Casey2015,Ma2015}.

In the 11$\arcmin \times$11$\arcmin$ deep medium bandwidth NIR survey area presented by \citet{Spitler2012} and \citet{AllenR2015}, the cluster shows four prominent cores that are linked by a filamentary structure.
The spatial metallicity distribution of the galaxies in the ZFIRE cluster indicates a few low-metallicity substructures that could be recently accreted on (T. Yuan; private communication).
Based on the spatial and redshift distribution traced by zCOSMOS SFGs and the population of DSFGs, it is possible that the large scale structure associated with the ZFIRE cluster may extend beyond the ZFIRE survey area.
In fact, \citet{Chiang2015} also detect three LAEs a few arcmins offset from the ZFIRE cluster (their positions are also shown in Figure~\ref{fig:dist_2dsky}).
Such large spatial extent traced by these various galaxies is consistent with the cosmological simulations \citep[e.g.,][]{Chiang2013}, where the $z\sim2.1$ cluster may still be actively accreting from large scale structure while more mature cluster cores begin to assemble in the densest regions.


\subsection{Significance of this structure}
In this section, we examine the significance of the large scale structure around the ZFIRE cluster traced by color-selected SFGs from zCOSMOS and {\it Herschel}-selected DSFGs.  
A total of 34 SFGs from the zCOSMOS survey is detected within a 10$\arcmin$ radius and a redshift slice of $\Delta z=$0.02.
We estimate the average number density of star-forming galaxies in the zCOSMOS survey first by excluding the data at $z=2.095\pm0.020$ and then smoothing the galaxy redshift distribution with $\Delta z=$0.1.
After interpolating the redshift distribution to $z=2.095$ and scaling to an area with 10$\arcmin$ radius and $\Delta z=$0.02, the expected galaxy number density is $\sim$8.9. 
This leads to an overdensity of zCOSMOS SFGs ($\delta_{\rm SFG}$) of $(34-8.9)/8.9=2.8^{+0.8}_{-0.7}$.

In Section 3.1, we show that a total of 8 DSFGs (excluding the one at $z=2.076$) are detected within a 10$\arcmin$ radius and a redshift slice of $\Delta z=$0.02.
The average number density of {\it Herschel}-selected galaxies with \LIR\ $\geq$10$^{12}$ \Lsun\ in the COSMOS field is $\sim$0.011 Mpc$^{-3}$ at $z\sim2.1$ \citep{LeeN2013}, in which we utilize the same procedures to select DSFGs and derive \LIR\ as described in Section 2.
Within a redshift slice of $\Delta z=$0.02 at $z\sim2.1$ and a $\sim$10\arcmin\ radius, the expected number of DSFGs is $\sim$0.6.
Using this expected number and assuming an uniformly distributed field, the probability of detecting 8 or more DSFGs within a 10\arcmin\ radius and $\Delta z=$0.02 with Poisson sampling is $<3\times10^{-7}$, suggesting that the excess of DSFGs is highly unlikely drawn by random chance.
A total of 8 DSFG within and around the ZFIRE cluster therefore corresponds to an overdensity $\delta_{\rm DSFG}=(8-0.6)/0.6=12.3^{+6.6}_{-4.6}$.

Assuming linear biasing, the overdensity of galaxies ($\delta_{\rm gal}$) can be related to the overdensity of mass ($\delta_{\rm m}$) with a known galaxy bias ($b$) and corrections of redshift space distortions from peculiar velocities ($C$), $1+b\delta_{\rm m}=C(1+\delta_{\rm gal})$ \citep{Steidel1998,Steidel2005}, where $C\equiv V_{\rm apparent}/V_{\rm true} =1+f-f(1+\delta_{\rm m})^{1/3}$ with $f=\Omega_{\rm M}(z)^{0.6}$.
We assume the bias of DSFGs is $\sim3.9$ based on the clustering analysis of {\it Herschel}-selected galaxies \citep{Mitchell-Wynne2012}, which leads to $\delta_{\rm m}=1.81^{+0.46}_{-0.56}$ as traced by DSFGs.
The errors are propagated from the uncertainties in $\delta_{\rm DSFG}$.
Similarly, assuming a galaxy bias of $\sim2$ for $BzK$-selected galaxies with comparable stellar masses (\Mstar) and SFR as the zCOSMOS SFGs in the structure \citep{Lin2012}, the overdensity of SFGs $\delta_{\rm SFG}=2.8^{+0.8}_{-0.6}$ leads to $\delta_{\rm m}=0.95^{+0.15}_{-0.19}$.
The mass overdensities $\delta_{\rm m}=1.81$ and $\delta_{\rm m}=0.95$ traced by DSFGs and SFGs correspond to linear overdensities of $\delta_{\rm L}=0.77$ and $\delta_{\rm L}=0.55$ in a spherical collapse model \citep{Mo1996}, in which they are expected to evolve to $\delta_{\rm L}\sim2$ and $\delta_{\rm L}\sim1.5$ at $z=0$.
The structure traced by DSFGs exceeds the collapse threshold ($\delta_{\rm c}=1.69$), and thus it is expected to virialize by $z=0$.
Yet SFGs with $\delta_{\rm L}\sim1.5$ implies that the structure does not collapse by $z=0$.

The estimate based on DSFG and SFG overdensities leads to inconsistent fate of the large scale structure around the ZFIRE cluster, although we note that the results based on SFGs and DSFGs are consistent with each other once taking into account the uncertainties in $\delta_{\rm SFG}$ and $\delta_{\rm DSFG}$.
\citet{Chapman2009} have reached a similar inconsistency in a $z=1.99$ structure, in which the overdensity traced by SMGs suggests that the structure can collapse whereas the overdensity traced by UV-selected galaxies suggests otherwise.
They thus conclude that SMGs must be even more biased tracer of mass than they assumed.
However, it is worth noting that a considerable number of assumptions and approximations have gone into these estimates based on the linear theory of spherical collapse model, yet the propagation of these uncertainties is not straightforward to determine.
In fact, \citet{Chiang2013} demonstrate that at $z\sim2$, $\delta_{\rm gal}$ of 2.6 for bright galaxies with \Mstar\ $>10^{10}$\Msun\ in a 25 Mpc (comoving) window is sufficient to collapse to Virgo-type clusters, suggesting that the overdensity traced by SFGs ($\delta_{\rm SFG}=2.8$) is likely to collapse by $z=0$.

\begin{figure}
\centering
  \includegraphics[width=0.5\textwidth]{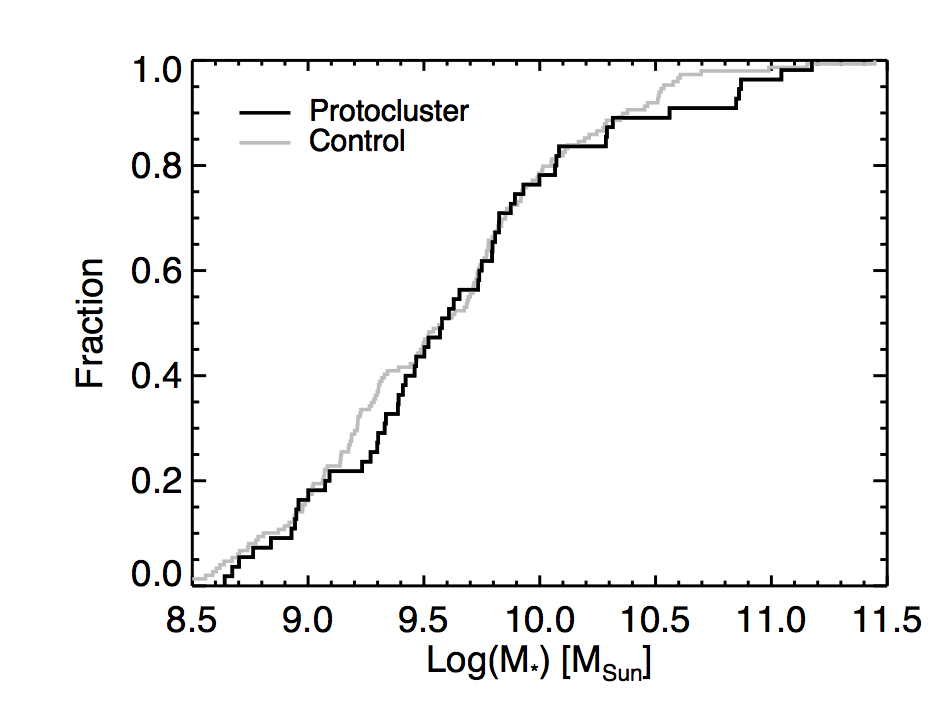} 
  \includegraphics[width=0.5\textwidth]{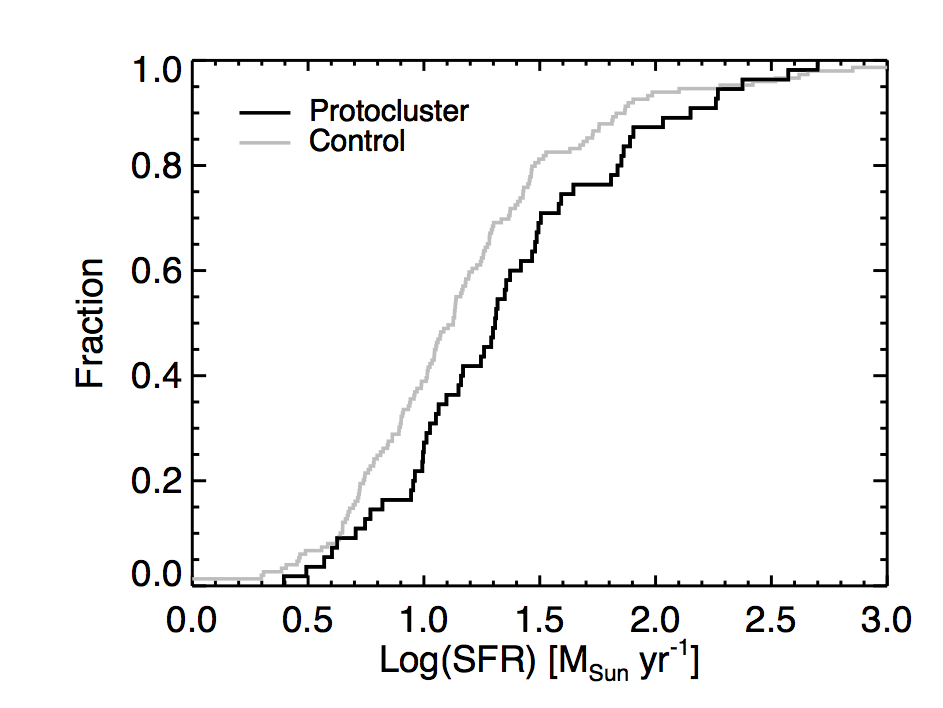} 
\caption{Normalized cumulative distributions of \Mstar\ (the top panel) and SFR (the bottom panel) of all SFGs from the zCOSMOS and SPIRE/\scuba\ surveys.
The black solid lines show the distributions of SFGs that can be linked to the ZFIRE cluster and the gray solid lines show the distributions of the control sample.
} 
\label{fig:dist_function}
\end{figure}

\section{Environmental Dependence}
The ZFIRE cluster \citep[][Y14]{Spitler2012} and its associated large scale structure traced by {\it Herschel}-selected DSFGs and color-selected SFGs represent a site that the cluster assembly may be actively taking place.
It is thus an ideal laboratory to test whether any early environmental impact have begun to influence galaxy evolution.
In this section, we examine the environmental dependence on galaxies' physical properties: \Mstar, SFR, gas content, and morphology.

\subsection{\Mstar\ and SFR}
We measure the \Mstar\ and SFR of all spectroscopically confirmed galaxies by fitting the full SEDs using the \textsc{Magphys} code \citep{da-Cunha2008}, and the HIGHZ extension \citep{da-Cunha2015} is used here since it includes stellar and dust emission priors that are more comparable to $z>1$ galaxies.
Since SED fitting-based SFRs often suffer from large uncertainties for highly obscured sources \citep[e.g.,][]{Wuyts2011}, we convert \LIR\ of DSFGs to their SFRs using the relation SFR/\Msun\,yr$^{-1}$=9.5$\times$10$^{-11}$ \LIR/\Lsun\ \citep[][adjusted for a Chabrier initial mass function]{Kennicutt1998a}.
The zCOSMOS SFGs (galaxies from redshift survey 3) that are linked to the ZFIRE cluster have mean \Mstar\ $=(1.23\pm0.36)\times10^{10}$ \Msun\ and mean SFR$=47\pm12$ \Msun\,yr$^{-1}$, where the uncertainties correspond to the standard deviation of the mean.
For the control sample, we use those SFGs from the same redshift survey (zCOSMOS; redshift survey 3) at the same redshift range ($2.07\leq z \leq 2.12$) but are not linked to the ZFIRE cluster based on the friends-of-friends analysis in Section 3.1.
This selection yields 140 SFGs with their mean \Mstar\ $=(7.17\pm0.80)\times10^{9}$ \Msun\ and mean SFR $=34\pm10$ \Msun\,yr$^{-1}$.


Figure~\ref{fig:dist_function} takes a closer examination of the \Mstar\ and SFR distributions of all SFGs in the large scale structure and the control sample (here we include all SFGs from redshift surveys 2 and 3).
A Kolmogorov-Smirnov (K-S) test does not reject either consistent or different distributions of \Mstar\ between the protocluster and the field (the p-value is 0.61).
However, the lack of an excess of massive galaxies in the protocluster environment may be a bias introduced by primarily selecting star-forming populations \citep[e.g.,][]{Muldrew2015}.
Despite protocluster and field SFGs have similar mean SFR, their SFR cumulative distributions show slight discrepancy, in which the cluster members are composed with slightly more high SFR galaxies but less low SFR galaxies (the p-value from the K-S test is 0.04).
This excess of high SFR galaxies in the large scale structure is consistent with the significant overdensity of DSFGs found within and around the ZFIRE cluster.

\subsection{Gas content}
One possible explanation for the excess of DSFGs in the protocluster is the enhanced gas supply in the dense environments at $z\gtrsim2$ \citep[e.g.,][]{Dave2010a}. 
Although no direct CO measurements are available for these SFGs, \citet{Scoville2014,Scoville2015} have developed an alternative probe of gas content of high$-z$ galaxies based on long wavelength dust continuum ($\lambda_{rest} \gtrsim 250 \mu$m).
Part of this large scale structure is covered by the \scuba\ 850 $\mu$m imaging survey \citep[with the root-mean-square noise down to $\sigma=0.8$ mJy,][]{Casey2013}, and thus we can constrain the gas content of protocluster and field SFGs via stacking analysis.
We note that here we enlarge our control sample defined in Section 4.1 by including both zCOSMOS SFGs at $2.07<z<2.12$ that are not linked to the ZFIRE cluster and SFGs at $2.04<z<2.07$ and $2.12<z<2.15$.

Following the stacking analysis outlined in \citet{Coppin2015}, we first remove 850 $\mu$m detected sources ($>3\sigma$) from \citet{Casey2013}.
These sources are removed by subtracting the image with PSFs at the source peak positions, and the PSFs are scaled according to the flux density of each source.
We measure the stacked flux densities of zCOSMOS SFGs in the protocluster and the control sample using \textsc{Simstack} \citep{Viero2013}, in which the flux densities are determined through regression with source-subtracted maps.
The measured $S_{850}$ of the zCOSMOS SFGs and the control sample are $-0.10\pm0.19$ mJy (25 sources) and $0.28\pm0.15$ mJy (43 sources), respectively.
The errors are propagated from the noise of individual stacked positions.
$S_{850}$ of the protocluster SFGs remains undetected even including the ZFIRE cluster members ($S_{850}=0.01\pm0.15$ mJy based on 47 sources). 
Our stacking sensitivity limit of $\sim0.15$ mJy corresponds to molecular gas mass ($M_{\rm mol}$) of $\sim1.5\times10^{10}$ \Msun\ \citep[based on the empirical calibration derived by][]{Scoville2015}, which leads to an upper limit of molecular gas fraction of $\sim$0.5 for the protocluster SFGs.

Of the SFGs that are included for stacking analysis, the mean SFR of zCOSMOS SFGs are comparable to the control sample.
The stacking results suggest that on average, the protocluster SFGs do not show enhanced gas supply compared to the field galaxies with comparable SFRs, and it is possible that their gas content may be even lower than the field galaxies.
These results are contradictory to the finding in the DSFG-rich $z=2.47$ protocluster \citep{Casey2015}, in which they show that the gas content of protocluster galaxies may be enhanced based on the same stacking analysis.
However, we stress that the $S_{\rm 850}$ detection of protocluster SFGs in \citet{Casey2015} is at 1.5$\sigma$ significance, and the detection of the control sample in this $z\sim2.1$ structure is at 2$\sigma$ significance.
Incorporating additional 850 $\mu$m observations \citep{Coppin2015,Koprowski2016} or follow-up molecular gas observations are needed to confirm these tentative trends.
With only a modest amount of Atacama Large Millimeter Array (ALMA) time (a few hours), the sensitivity of $S_{850}$ can be improved by a factor of 10 when stacking $\sim$30 sources.

\subsection{Galaxy morphology}
Part of this $z\sim2.1$ structure is covered by the deep WFC3 F125W/F160W images from CANDELS \citep[Cosmic Assembly Near-infrared Deep Extragalactic Legacy Survey;][]{Grogin2011,Koekemoer2011}, enabling our examination of the rest-frame optical morphology for a subset of SFGs in the protocluster and the control sample.
\citet{AllenR2015} show that there is no difference of average \sersic\ index of SFGs in the ZFIRE cluster and the control sample from ZFOURGE.
Here we examine if this result extends toward zCOSMOS SFGs which trace the structure within and beyond the ZFIRE cluster.
Based on a catalog of galaxy structural parameters from \citet{van-der-Wel2012,van-der-Wel2014}, we find that the mean \sersic\ index  of the zCOSMOS SFGs (based on 18 galaxies with reliable structural parameters) is $1.64\pm0.23$, which is consistent with the mean sersic index of the cluster star-forming galaxies in \citet{AllenR2015}.

In addition to the morphological properties based on parametric fits, we use an $H$-band morphology catalog based on neural networks from \citet{Huertas-Company2015}, in which the algorithm is trained based on the visual classifications in CANDELS GOOD-S field \citep{Kartaltepe2015}.
In the catalog, each galaxy is assigned the probabilities of having a spheroid, a disk, showing irregular feature, being a point source, and being unclassifiable.
Based on these morphological classifications, we find that the zCOSMOS SFGs show higher spheroid fractions but similar disk and irregular fractions compared to the control sample.
For the zCOSMOS SFGs (based on 26 galaxies), the fraction of galaxies with a prominent spheroid component (probability of spheroid, $f_{\rm sph}>0.67$), a prominent disk component (probability of disk, $f_{\rm disk}>0.67$), and obvious irregular features (probability of irregular, $f_{\rm irr}>0.67$) are $0.28^{+0.15}_{-0.10}$, $0.44^{+0.18}_{-0.13}$, and $0.24^{+0.14}_{-0.09}$, respectively.
In the control sample (based on 28 galaxies), these spheroid, disk, and irregular fractions are $0.07^{+0.10}_{-0.05}$, $0.55^{+0.18}_{-0.14}$, and $0.15^{+0.12}_{-0.07}$, respectively.
These differences are suggestive that SFGs in this large scale structure may be undergoing more violent assembly through rapid accretion or galaxy interaction, where these processes can lead to fast build-up of bulge \citep[e.g.,][]{Hopkins2006a,Ceverino2010}. 

Several studies at $z\sim1$ have demonstrated an increase of galaxy merger fractions in dense environments \citep[e.g.,][]{Lin2010,Sobral2011,de-Ravel2011}.
Here we examine if the frequency of galaxy mergers/interacting systems are indeed higher in the protocluster using visual inspection.
We assign galaxies as merger candidates if they show prominent irregular features and/or having a close companion \citep[e.g.,][]{Kartaltepe2012,Hung2013}.
The merger fractions of zCOSMOS SFGs and the control sample are $0.46^{+0.17}_{-0.13}$ and $0.36^{+0.15}_{-0.11}$, respectively.
These fractions are similar if we define merger candidates as galaxies with irregular fractions of $f_{\rm irr}>0.4$ using the \citet{Huertas-Company2015} classifications.
No significant differences in galaxy merger fractions are measured between zCOSMOS SFGs and the control sample.

\begin{figure}
\centering
  \includegraphics[width=0.45\textwidth]{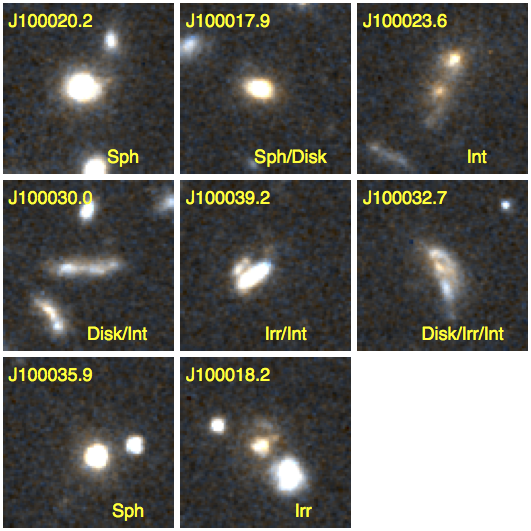} 
\caption{Three-color images of 8 DSFGs in the $z\sim2.1$ structure with CANDELS WFC3 images available.  
WFC3 F125W, (F125W+F160W)/2 and, F160W used for the blue, green and, red
channel, respectively. 
All images have the size of $5\arcsec \times 5\arcsec$.
The short name and morphology type of each DSFG are displayed in each panel.
``Sph'', ``Disk'', and ``Irr'' are shown when $f_{\rm sph}$, $f_{\rm disk}$, and $f_{\rm irr}$ are $>0.5$.
DSFGs indicated as ``Int'' are candidates of mergers/interacting galaxies.
} 
\label{fig:wfc3panel}
\end{figure}

\section{Triggering of DSFGs in this structure}
One important question to address is the origin of excess DSFGs in this $z\sim2.1$ structure (both in the ZFIRE cluster and its associated large scale structure).
A possible explanation is that the underlying star-forming galaxy main sequence \citep[e.g.,][]{Brinchmann2004,Noeske2007} in the protocluster environments are intrinsically different from the field.
An elevated galaxy main sequence can naturally lead to an increase of high SFR population at a given \Mstar.
However, \citet{Koyama2013} have shown that the SFR-\Mstar\ relation of star-forming galaxies is independent of environments at $z\sim2$, and in this $z\sim2.1$ structure, the mean specific SFR of protocluster SFGs does not differ significantly from the control sample.
We thus conclude that there is no obvious evidence to attribute the excess DSFGs to an elevated galaxy main sequence.

The excess of DSFGs in the protoclusters represents an enhanced population of galaxies that are ``outliers'' of the SFR-\Mstar\ relation at $z\sim2$, and these DSFGs may be triggered by an enhanced gas supply (with respect to the SFGs on the galaxy main sequence) or through galaxy interactions \citep[e.g.,][]{Tacconi2010,Engel2010}.
However, based on the stacking and morphological analysis in Section 4.2 and 4.3, we find no evidence that the protocluster SFGs exhibit more gas supply or higher merger fraction than the field.
Among the 8 DSFGs that have CANDELS WFC3 images available (Figure~\ref{fig:wfc3panel}), four of them may exhibit ongoing interaction (J100031.8, J100030.0, J100039.2, and J100032.7).
This leads to a merger fraction of 0.5, which is also consistent with the field DSFGs at $z\sim2$ \citep{Kartaltepe2012}.

\section{Discussion and Summary}
We present a search of large scale structure around a $z=2.095$ cluster \citep[][Y14]{Spitler2012}.
Within a 10$\arcmin$ (corresponds to a proper distance of $\sim5$ Mpc) radius from the cluster center and a redshift range of $2.07\leq z \leq 2.12$, there are 9 DSFGs (including 4 X-ray luminous AGNs), and 34 $BzK$- and UV-selected SFGs.
This leads to galaxy overdensities of $\delta_{\rm DSFG}\sim12.3$ and $\delta_{\rm SFG}\sim2.8$.
An estimation based on the linear theory of spherical collapse model suggests that only the overdensity traced by DSFGs can collapse by $z\sim0$.
However, $\delta_{\rm SFG}$ of 2.8 is only slightly under the collapsing threshold, and it is expected to collapse to a Virgo-type structure when comparing to the prediction of cosmological simulations with \Mstar\ $>10^{10}$ \Msun\ in a 25 (comoving) Mpc window \citep{Chiang2013}.
The ZFIRE cluster and its associated DSFG- and SFG-rich structure represents an active cluster formation phase, in which the $z\sim2.1$ cluster is still accreting from large scale structure while more mature cluster cores begin to assemble.
This structure is thus an ideal site to explore early environmental dependence of galaxy properties, and the interplay between intergalactic medium and galaxies during the formation of massive clusters \citep[e.g.,][]{LeeKG2014,LeeKG2016,Cai2015}.

The identification of this $z\sim2.1$ structure, along with previous finding of DSFG- or AGN-rich structure in {\it known} protoclusters \citep{Tamura2009,Digby-North2010,Smail2014,Dannerbauer2014}, demonstrate that the excess of these rare systems indeed trace bona-fide overdensities (of mass) and protoclusters \citep[also see][]{Casey2015}.
In fact, searching for the overdensities of DSFGs or AGNs can be an efficient probe\footnote{The sample of protoclusters identified this way may be biased, since only a fraction of $z\gtrsim2$ protoclusters are expected to host excess DSFGs \citep{Casey2016}.} of large structure across several tens Mpc at $z\gtrsim2$, and there is potentially a large population of such DSFG-rich protoclusters as revealed by {\it Herschel} and {\it Planck} observations \citep{Planck-Collaboration2015}.
Without prior knowledge of the ZFIRE cluster, it is still possible to identify this structure relying only on the excess of DSFGs.
The redshift survey targeted on SPIRE/\scuba-bright sources finds 4 DSFGs within an area of $\sim140$ squared arcmin on the sky and $\Delta z=0.02$ (orange dots with red circle in Figure~\ref{fig:dist_2dsky}), which leads to an overdensity of $\delta_{\rm DSFG}\sim10.4$.

A clear picture that explains the triggering of excess DSFGs/AGNs has not yet emerged.
It is plausible that galaxies in dense environments at $z\gtrsim2$ exhibit elevated gas supply and higher chances of galaxy interactions.
However, our morphological analysis shows that the merger fraction of the protocluster galaxies is consistent with the control sample, and there is also no obvious enhancements of mergers/interacting systems in protocluster DSFGs as compared to the field populations.
Follow-up high resolution observations with integral field spectrographs can provide further information on the dynamics and/or reveal recent merger histories of these DSFGs \citep[e.g.,][]{Hung2016}. 
The stacking results based on 850 $\mu$m images are not yet sensitive enough to constrain the gas content of both protocluster and field galaxies, and it is necessary to incorporate additional submillimeter imaging or pursue follow-up molecular gas observations for both SFGs and DSFGs.

\acknowledgments
C-LH thanks J. R. Mart{\'{\i}}nez-Galarza for helpful discussion on SED fitting.
C-LH acknowledges support from the Harlan J. Smith Fellowship.
CMC thanks the University of Texas at Austin, College of Natural Science for support.
GGK acknowledges the support of the Australian Research Council through the award of a Future Fellowship (FT140100933).

The authors wish to recognize and acknowledge the very significant cultural role and reverence that the summit of Mauna Kea has always had within the indigenous Hawaiian community.  
We are most fortunate to have the opportunity to conduct observations from this mountain.
Some of the data presented here were obtained at the W. M. Keck Observatory, which is operated as a scientific partnership among the California Institute of Technology, the University of California, and the National Aeronautics and Space Administration and made possible by financial support of the W. M. Keck Foundation. 

Part of the data presented in this paper has been made available by the COSMOS team (http://cosmos.astro.caltech.edu).
COSMOS is based on observations with the NASA/ESA Hubble Space Telescope, obtained at the Space Telescope Science Institute, which is operated by AURA Inc, under NASA contract NAS 5-26555; also based on data collected at : the Subaru Telescope, which is operated by the National Astronomical Observatory of Japan; the XMM-Newton, an ESA science mission with instruments and contributions directly funded by ESA Member States and NASA; the European Southern Observatory, Chile; Kitt Peak National Observatory, Cerro Tololo Inter-American Observatory, and the National Optical Astronomy Observatory, which are operated by the Association of Universities for Research in Astronomy, Inc. (AURA) under cooperative agreement with the National Science Foundation; the National Radio Astronomy Observatory which is a facility of the National Science Foundation operated under cooperative agreement by Associated Universities, Inc ; and the Canada-France-Hawaii Telescope operated by the National Research Council of Canada, the Centre National de la Recherche Scientifique de France and the University of Hawaii.


\bibliographystyle{../Reference_Bib/apj}
\bibliography{../Reference_Bib/Cosmos}

\end{document}